\documentclass[12pt,a4paper]{article}
\usepackage{amssymb,amsmath,graphicx,subcaption,cite}

\usepackage{epstopdf}
\usepackage[utf8]{inputenc}
\usepackage[english]{babel}

\begin{document}

	
\begin{center}
   {\Large\bf  Rodlike Heisenberg nanomagnet driven by propagating magnetic field: Nonequilibrium phase transition}
\end{center}
\vskip 1 cm
\begin{center} 
    Muktish Acharyya$^{*}$
   
   \textit{Department of Physics, Presidency University,}\\
   \textit{86/1 College Street, Kolkata-700073, India} 
\vskip 0.2 cm
   
   {Email$^*$:muktish.physics@presiuniv.ac.in}
\end{center}
\vspace {1.0 cm}
	
\noindent {\bf Abstract:} The dynamical responses of a rodlike anisotropic Heisenberg ferromagnet, irradiated by
the propagating magnetic field wave, have been studied by Monte Carlo simulation using Metropolis
single spin-flip algorithm. A nonequilibrium dynamical phase transition has been observed. The
transition temperature has been obtained from the peak position of the variance of the dynamic
order parameter plotted as a function of the temperature. The transition has been found to occur at a lower temperature for the higher value
of the amplitude of the propagating magnetic field wave. A comprehensive phase boundary was
drawn. The phase boundary has been found to be insensitive to the wavelength of the propagating
field wave. The boundary encloses more area of the region of the ordered phase for stronger
anisotropy.
\vskip 4cm

\textbf{Keywords: Heisenberg ferromagnet, Monte Carlo simulation, Metropolis 
algorithm, Magnetic anisotropy}

\vskip 1cm

\newpage
\noindent{\large\bf I. Introduction}

The studies on magnets have taken important places in modern research over several decades due to various interesting physical phenomena and commercial applications. The study of
the magnetic materials in the nanometer scale has drawn additional attention to the modern
research\cite{rana,irshad}. The most important applications of such nanomagnetic devices, in the field of magnetic storage, are mainly due to the advanced technological features\cite{barman}. The basic challenge is to develop a nanomagnetic device with a very high density of storage.

The theoretical research, to study the behaviors of tiny magnetic materials, is also
advancing in parallel to the experimental research. The dynamics of Ising like core-shell nanoparticle \cite{erol2,kantar1} and the dynamical responses of uniaxial
nanomagnetic system driven by an oscillating magnetic field are recently studied
\cite{yuksel1}. The simple Ising
type interaction was considered to have the response of such kinds of systems on a small
length scale. The uniaxial nanowire has been investigated\cite{yuksel2}
to study the nonequilibrium
phase transition.

The rodlike nanomagnets have also been investigated recently. The cylindrical magnetic nanowire driven by propagating magnetic field wave has been investigated\cite{erol1} recently with extensive Monte Carlo simulation to get interesting nonequilibrium phase transition.
The results of magneto-kinetic properties have been reported\cite{deviren2} in
cylindrical Ising nanotube. The dynamic responses of hexagonal Ising nanowire have
been investigated\cite{ertas1}. The cylindrical Ising nanowire, driven by an oscillating magnetic field, has shown\cite{keskin1} a nonequilibrium phase transition.

All the simulational studies mentioned above are mainly based on the Ising type interaction
among the magnetic spins. This Ising type interaction could be the best model of highly
anisotropic magnetic systems. The phase boundary of the nonequilibrium phase transition
can depend on the anisotropy of the magnetic system. Such kind of variation cannot be observed in the case of Ising-like interaction. In this context, the most general classical spin model is the anisotropic
Heisenberg model. In this paper, the classical anisotropic Heisenberg model is used
to study the dynamical responses of a rodlike nanomagnet irradiated by propagating
magnetic field waves. This will serve as the general model to study the response of rodlike nanomagnet to the optical perturbations.

The manuscript is organized as follows: the next section (section-II) is devoted to introducing the
model, section-III describes the simulational scheme, the numerical results are reported in section-IV, the paper ends with a summary in
section-V.

\vspace {0.5 cm}

\vskip 0.5 cm
\noindent {\large\bf II. The Model of rodlike Heisenberg nanoferromagnet:}
\vskip 0.5 cm

Here, the system of interest is the anisotropic classical Heisenberg ferromagnet. It can be represented by the following Hamiltonian

\begin{equation}
H= -J\sum_{<ij>} {\vec S_i} \cdot {\vec S_j} - D\sum_i (S_i^z)^2 - \sum {\vec H(i,j,t)} \cdot {\vec S_i}
\end{equation}

\noindent where the ${\vec S_i}$ represents the Heisenberg spin vector ($|{\vec S_i}|=1$) at i-th site, having three cartesian components $(S_i^x, S_i^y, S_i^z)$. The first term in the above expression represents the contribution to the energy arises due to spin-spin Heisenberg nearest neighbour ($<ij>$) ferromagnetic ($J > 0$) interaction.
The second term is the single-site anisotropy (crystal field with z-axis as easy
axis) and the third is the Zeeman term where the interaction of spin with externally
applied space-time dependent magnetic field ${\vec H(i,j,t)}$ (measured in the unit of $J$). The space-time
dependent magnetic field is propagating along the z-direction (represented by the index k of the three-dimensional lattice) and
is polarized in the x-direction with the following mathematical form:
\begin{equation}
H(i,j,t) = h_x {\rm cos}[2\pi(ft-k/{\lambda})]
\end{equation}
\noindent where $h_x$ is the amplitude, $f$ is the frequency and $\lambda$ is the
wavelength of the propagating magnetic field wave.

Here, the propagating magnetic field is considered as the magnetic component of the propagating electromagnetic wave, keeping in mind, to study the response of Heisenberg nanomagnetic rod irradiated by an electromagnetic field. Here, the magnetic polarisation of the transverse EM wave (propagating in the z-direction) is considered in the x-direction.

The geometry of the system is quasi one dimensional of size $N=L_x\times L_y \times L_z$ where $L_x=L_y=4$ and $L_z= 200$. This describes the structure of rodlike
Heisenberg nanomagnet. The magnetic wave is propagating in the z-direction.
The system is exposed to the open boundary conditions in all directions.

\vskip 2 cm
\noindent {\large\bf III. Monte Carlo methods applied to Heisenberg nanomagnet:} 
\vskip 0.5 cm

The system is initially kept at a very high temperature, guaranteed by the random
spin configuration of the paramagnetic phase. Then the magnetic field is
applied. The magnetic field (space-time dependent) is propagating along the
z-direction (longitudinal axis of the system considered here). The space-time-dependent
magnetic field keeps the system far away from the equilibrium. However, the
nonequilibrium steady state of the system can be achieved at any lower temperature
by cooling the system slowly from the initial high-temperature spin configuration.

The spins of the system are updated by a random updating scheme with Metropolis
single spin-flip algorithm. A lattice site is chosen randomly. The initial spin state
of that site is noted. A trial spin configuration of that chosen site is chosen
randomly. The change in energy ($\Delta H$) has been calculated from the given
form of the Hamiltonian (equation-1) with nearest-neighbor interactions. This
trial spin configuration will be accepted with Metropolis probability\cite{binder,newman}

\begin{equation}
P({\vec S}_i^{initial} \to {\vec S}_i^{final}) = {\rm Min}[1, e^{({{-\Delta H} \over {kT}})}]
\end{equation}

The trial spin configuration will be accepted if another random number
(uniformly distributed between 0 and 1) is less than or equal to the Metropolis
probability mentioned above (equation-2). In this way, $N$ number of such updates
have been done. $N$ such random updates constitute the time unit and are termed as
MCSS (Monte Carlo Step per Site). At any fixed temperature, $16\times10^5$ MCSS
are allowed. The initial $8\times10^5$ MCSS are discarded (as assumed the transient
period to achieve the nonequilibrium steady state (NESS)). The quantities are measured
from the averaging over the next $8\times10^5$ MCSS. The stepsize of the temperature for cooling the system
was considered $\Delta T=0.05$. The unit of temperature is $J/k$. This value of $\Delta
T$ is assumed to be small enough to move from one nonequilibrium steady state (NESS) to lower temperature NESS.

\vskip 1cm
\noindent {\Large{\bf IV. Simulational results:}}
\vskip 1cm

We are mainly interested to study the nonequilibrium phases depending on the
temperature of the system. In this context, the relevant measurable quantities
would be the component of the nonequilibrium dynamic order parameter, the variance of that component of the order
parameter (so-called susceptibility). Keeping these in mind, the following quantities are calculated:

\noindent (i) Components of instantaneous magnetisation
$m_{\alpha}(t)={{1} \over {N}}\sum S_i^{\alpha}$
where $\alpha$ denotes different Cartesian components, $\alpha = x,y,z$.

\noindent (ii) Components of time averaged magnetisation over a full cycle of the propagating magnetic field:
$q_{\alpha} = {{1} \over {f}}\int m_{\alpha}(t) dt$.
and
$Q_{\alpha} = <q_{\alpha}>$. The averaging, represented by $<...>$, has been calculated over various cycles of the propagating magnetic field. In this study, the frequency of the propagating magnetic wave is considered $f=0.01$. So the 100 MCSS will constitute a complete cycle of the wave. This quantity $Q_{\alpha}$ serves as the dynamical order parameter. It may be mentioned here that such a vector order parameter was introduced \cite{ma1, ma2} in the Monte Carlo study of Heisenberg ferromagnets and verified experimentally \cite{robb, berger1, berger2, berger3} afterwards.

\noindent (iii) The variance of the dynamic order parameter:

$\chi_{\alpha} = N ( <q^2_{\alpha}> - <q_{\alpha}>^2)$

The z-component of the dynamical order parameter has been investigated as a function of temperature (T) to study the nonequilibrium phase transition. The dynamical order parameter, $Q_z$, was observed to become nonzero at any finite temperature ($T_d$) if the system was cooled down from a high temperature ($Q_z=0$) for a fixed set of $\lambda$, $D$, and $h_x$ values. This indicates the nonequilibrium phase transition. The direction of polarisation of the magnetic field (of the propagating wave) and the easy anisotropy axis would make $Q_x=Q_y=0$. Here, the dynamic ordering is possible only in the z-direction.

\begin{figure}[h!]
\centering
	\begin{subfigure}{0.48\textwidth}
	\includegraphics[angle=0,width=\textwidth]{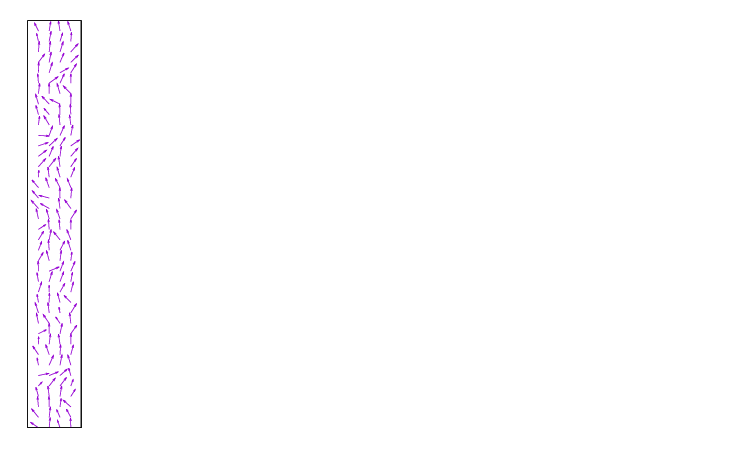}
	\subcaption{}
	\end{subfigure}
	\begin{subfigure}{0.48\textwidth}
	\includegraphics[angle=0,width=\textwidth]{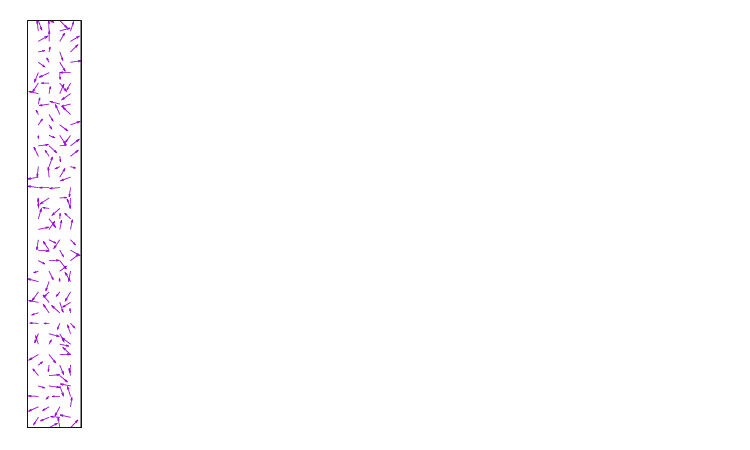}
	\subcaption{}
	\end{subfigure}

\caption{Longitudinal cross sections of Spin projections at $t=16000$ MCSS  (a) $T=0.60$ and (b) $T=2.40$. Here, $D=0.50$, $h_x=1.20$, $f=0.01$ and $\lambda=10$.
The direction of propagation is upward.}
\label{structure}
\end{figure}

The longitudinal cross-section of the spin projections is shown in Fig. -\ref{structure} for two different values of temperature at any instant of time, t = 16000 MCSS. The ordered structure is shown in the low temperature ($T=0.60$) configuration (Fig-\ref {structure}(a)), which provides the dynamically ordered phase.However, this ordered structure disappears at high temperatures ($T = 2.40 $) in a dynamically disordered phase (Fig. \ref {structure}(b)).

\begin{figure}[h!]
\centering
	\begin{subfigure}{0.45\textwidth}
	\includegraphics[angle=0,width=\textwidth]{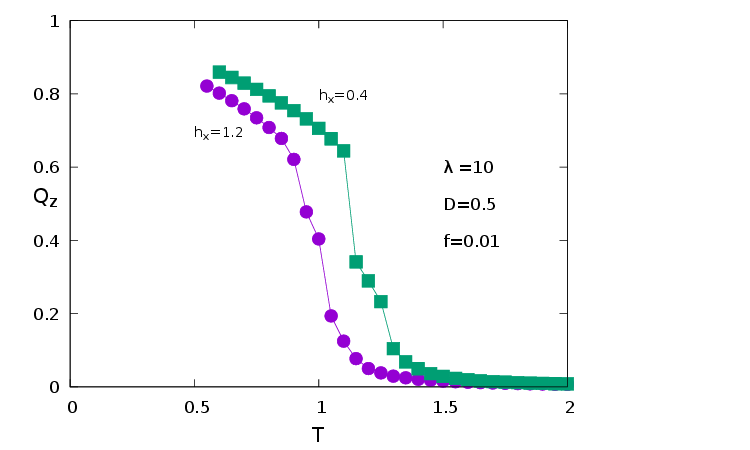}
	\subcaption{}
	\end{subfigure}
	\begin{subfigure}{0.45\textwidth}
	\includegraphics[angle=0,width=\textwidth]{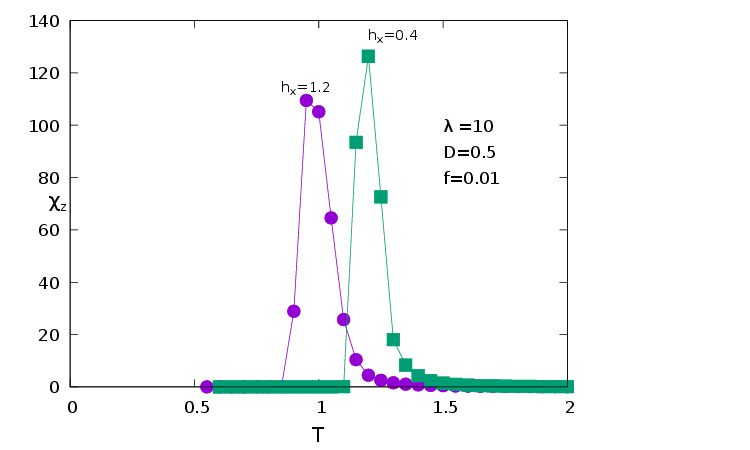}
	\subcaption{}
	\end{subfigure}
	\begin{subfigure}{0.45\textwidth}
	\includegraphics[angle=0,width=\textwidth]{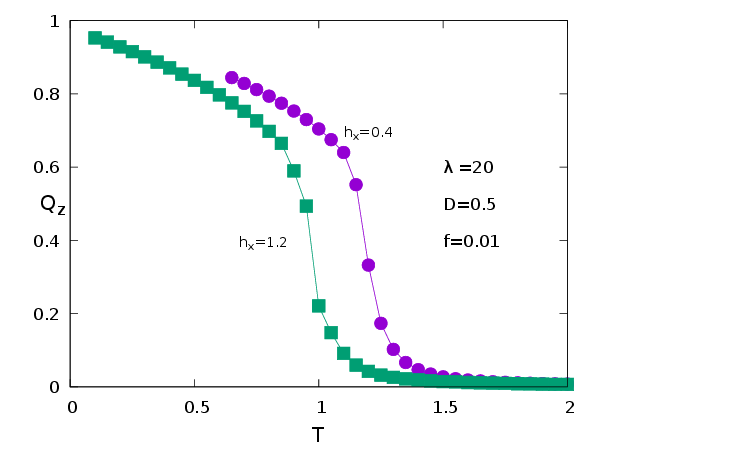}
	\subcaption{}
	\end{subfigure}
	\begin{subfigure}{0.45\textwidth}
	\includegraphics[angle=0,width=\textwidth]{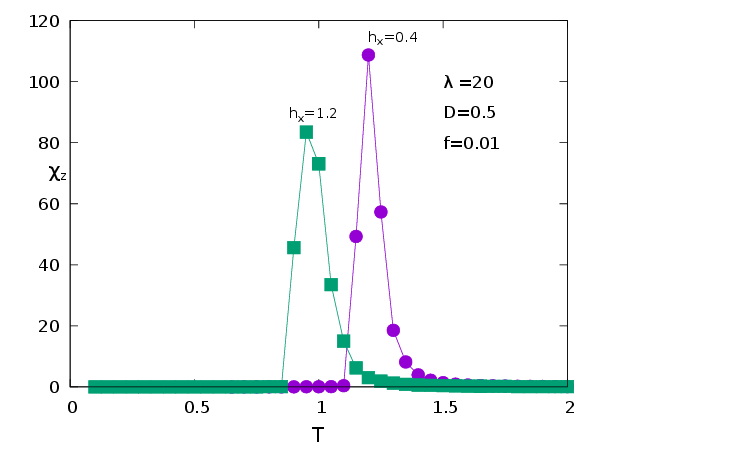}
	\subcaption{}
	\end{subfigure}

\caption{The order parameters ($Q_z$) and its variances ($\chi_z$) are plotted against the temperature ($T$) for $D=0.5$. Here, $L_z=200$.}
\label{ordD0.5}
\end{figure}

The transition temperature, for this nonequilibrium dynamical phase transition, is denoted as $T_d$. The transition occurs at a lower temperature for a higher value of applied field amplitude $h_x$. One such typical plot is shown in Fig-\ref{ordD0.5}(a). Here, we have used, $D=0.5$, $\lambda=10$. The frequency $f$ of the propagating magnetic field was kept fixed ($f=0.01$) throughout the study. The dynamical ordering takes place at higher temperature for $h_x=0.4$.

The  variance ($\chi_z$) of the dynamic order parameter ($Q_z$) was studied as a function of the temperature and is plotted in Fig-\ref{ordD0.5}(b). The variance was found to be peaked sharply (assumed to have diverged in thermodynamic limit) at the dynamic transition temperature $T_d$. For, $h_x=0.4$ the transition occurs at $T_d=1.2$ and for $h_x=1.2$ the transition occurs at $T=0.95$. In fact, to measure $T_d$, the position of the peak of the variance of the dynamic order parameter was used here. The variations, of the order parameter $Q_z$) and its variance $\chi$ near the transition temperature, are consistent with those observed generally in the case of equilibrium phase transitions. It may be noted here (as observed also) that the transition temperature $T_d$ depends on the amplitude ($h_x$) of the propagating magnetic field wave. It is found that the dynamic transition occurs at lower temperatures for higher values of $h_x$.

Do we expect any significant dependence of the transition temperature on the wavelength ($\lambda $) of the propagating field wave? To know the answer, we have studied this for a different value of $\lambda$ and shown the results in Fig-\ref{ordD0.5}(c) (the temperature dependences of the order parameter) and Fig-\ref{ordD0.5} (d) (the temperature dependences of the variance of the dynamic order parameter). No such significant changes in the dynamic transition temperature were observed in this case. That will be clearly shown in the phase diagram.

\begin{figure}[h!]
\centering
	\begin{subfigure}{0.45\textwidth}
	\includegraphics[angle=0,width=\textwidth]{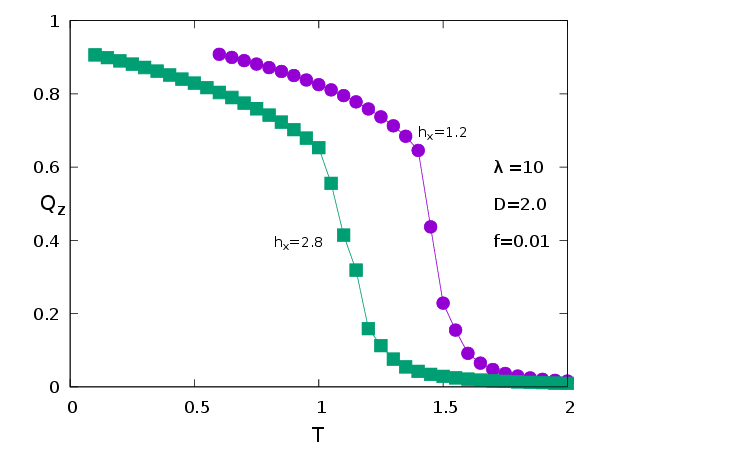}
	\subcaption{}
	\end{subfigure}
	\begin{subfigure}{0.45\textwidth}
	\includegraphics[angle=0,width=\textwidth]{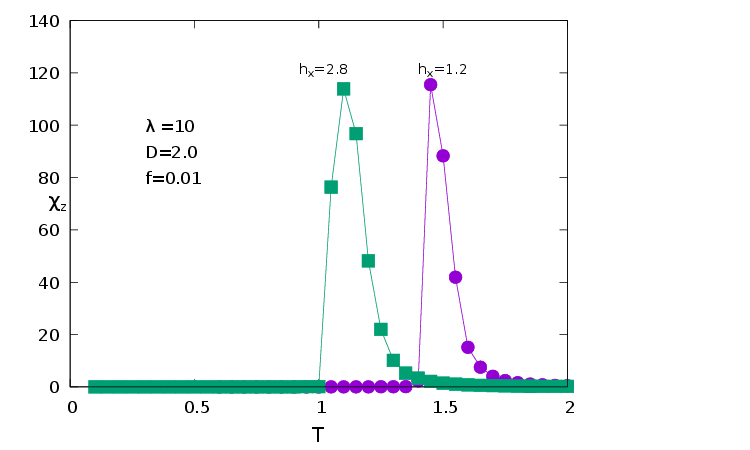}
	\subcaption{}
	\end{subfigure}
	\begin{subfigure}{0.45\textwidth}
	\includegraphics[angle=0,width=\textwidth]{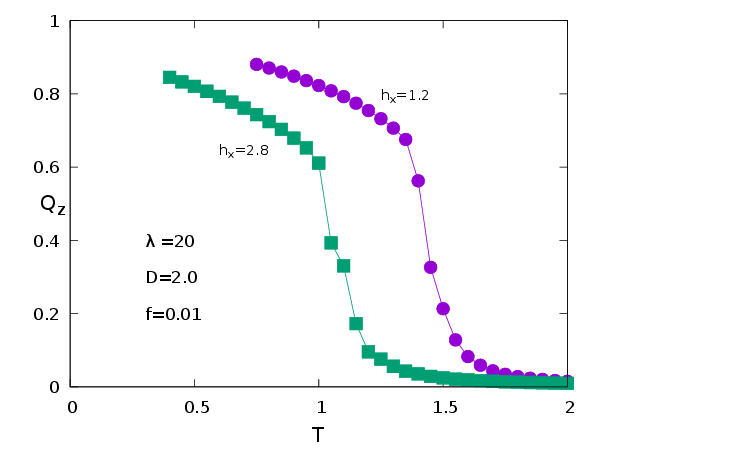}
	\subcaption{}
	\end{subfigure}
	\begin{subfigure}{0.45\textwidth}
	\includegraphics[angle=0,width=\textwidth]{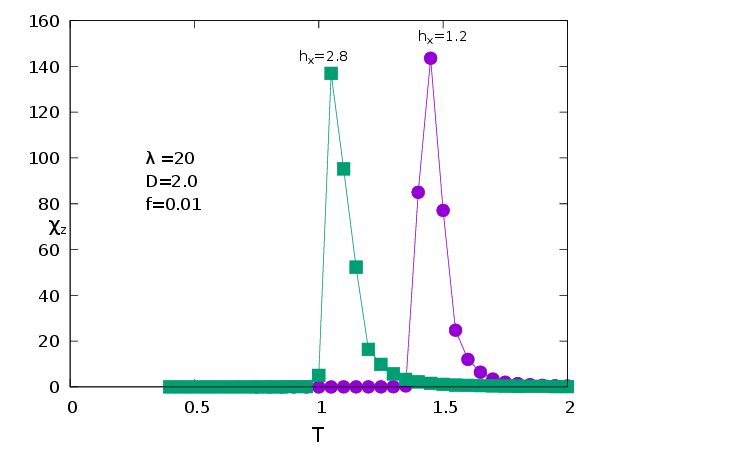}
	\subcaption{}
	\end{subfigure}
\caption{The order parameters ($Q_z$) and its variances ($\chi_z$) are plotted
against the temperature ($T$) for $D=2.0$. Here, $L_z=200$.}
\label{ordD2.0}
\end{figure}

However, we have observed significant changes in the dynamic transition temperature ($T_d$) as the strength of anisotropy is changed. The same studies mentioned above (and the results shown in Fig-\ref {ordD0.5}) are done for a different value of the strength of anisotropy, $D = 2.0 $. For other parameters held constant ($h_x =1.2$, $\lambda =10$, and $f =0.01$), the transition temperature was found to increase (as shown in Fig-\ref{ordD2.0}(a) and Fig-\ref{ordD2.0}(b)). In this case, the transition temperature was found at $T_d = 1.45$ (it was $T_d = 0.95$ for $h_x = 1.2 $, $D = 0.5$, and $\lambda = 10$ as shown in Fig.\ref{ordD0.5}(b)). The transition temperature did not change significantly when the wavelength of the propagating magnetic field wave ($\lambda =20$) was increased. These are shown in Fig. \ref{ordD2.0}(c) and Fig. \ref{ordD2.0}(d).

\begin{figure}[h!]
\centering
	\begin{subfigure}{0.45\textwidth}
	\includegraphics[angle=0,width=\textwidth]{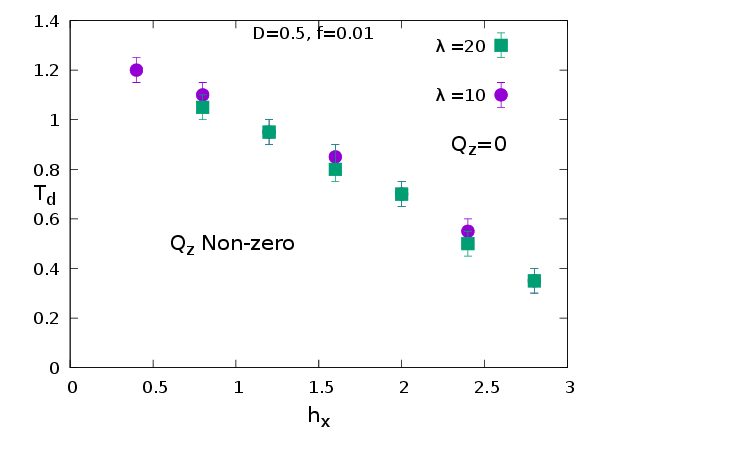}
	\subcaption{}
	\end{subfigure}
	\begin{subfigure}{0.45\textwidth}
	\includegraphics[angle=0,width=\textwidth]{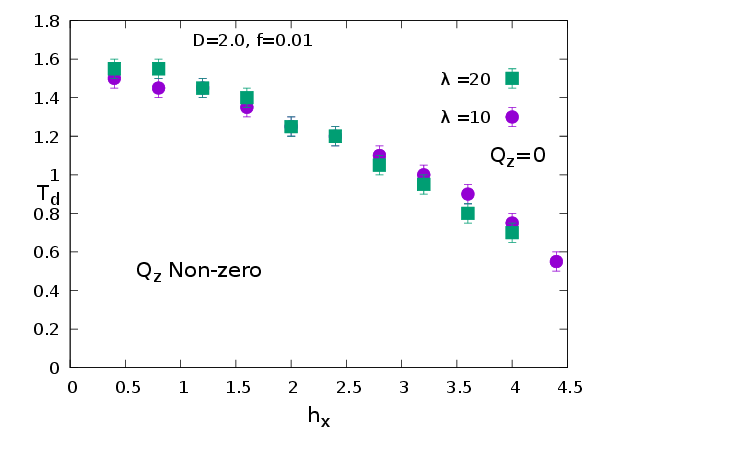}
	\subcaption{}
	\end{subfigure}
\caption{The phase diagrams of nonequilibrium phase transitions. (a) for D=0.5 and
(b) for D=2.0. Here $L_z=200.$}
\label{phase}
\end{figure}

A comprehensive phase boundary has been drawn on the $T_d-h_x$ plane with
$lambda$ and $D$ being different parameters. The peak position of the variance
($\chi_z$) of the
dynamic order parameter was used to estimate the dynamic transition
temperatures $T_d$ for different parameters. Fig-\ref{phase}(a) shows such a
phase diagram for $D=0.5$, $f=0.01$ and two different values of $\lambda$. The
region of non-zero (dynamically ordered phase) $Q_z$ and the region of $Q_z=0$ (dynamically disordered phase) are shown. The phase diagrams of two different
values of $\lambda$ (=10 and 20) are shown in different symbols (and with different
colors). The phase boundaries do not show any significant dependence on the
wavelengths of the propagating magnetic field waves. The same kinds of phase diagrams
are shown (in Fig-\ref{phase}(b)) for different values of the strength of anisotropy $D=2.0$. Here the significant variation of the phase boundary is observed. For
the higher value of the anisotropy, the region of the ordered phase (non-zero $Q_z$)
is expanded.

\begin{figure}[h!]
\centering
	\includegraphics[angle=0,width=\textwidth]{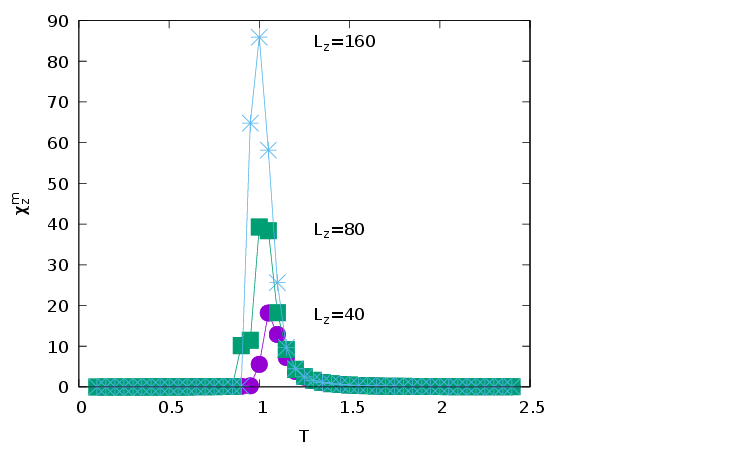}
\caption{The variances ($\chi_z$) of the order parameter ($Q_z$) are plotted against the
 temperature ($T$) for different longitudinal sizes ($L_z$) of the nanorod.Here, $D=0.5, \lambda=10, h_x=1.2$
 and  $f=0.01$.}
\label{growth}
\end{figure}

The variance ($\chi_z$) of the dynamic order parameter ($Q_z$) has been studied as
a function of temperature ($T$) for different values of $L_z$. The results are shown
in Fig-\ref{growth}. The height of the peak of $\chi_z$ has been observed to increase
as the length ($L_z$) of the nanorod increases. This is a sign of the growth of dynamic
correlations near the transition temperature. This will eventually diverge for $L_z \to \infty$ indicating the
true phase transition (of nonequilibrium kind) in the thermodynamic limit.

\begin{figure}[h!]
\centering
	\includegraphics[angle=0,width=\textwidth]{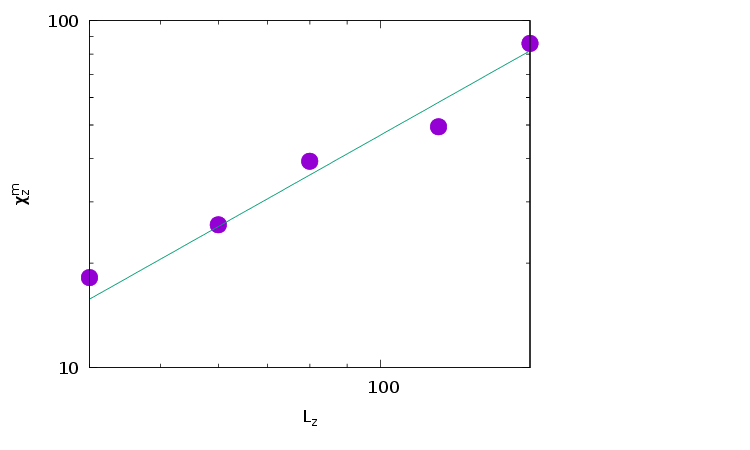}
\caption{The maxima ($\chi^m_z$) of the variances ($\chi_z$) of the order parameter ($Q_z$) have been plotted against the
 longitudinal
 sizes ($L_z$) of the nanorod
, in logarithmic scale. The solid line represent the fitting function $Y
=0.196X^{1.1887}$. Here, $D
=0.5, \lambda=10, h_x
=1.2$
 and $f=0.01$.}
\label{exponent}
\end{figure}

The peak values ($\chi_z^m$) of the variances ($\chi_z$) of the order parameter ($Q_z$)
for different longitudinal sizes ($L_z$=40,60,80,120 and 160 here) of the system are plotted in logarithmic
scale and shown in Fig-\ref{exponent}. The data points are fitted in a power law
$\chi^p_z \sim (L_z)^{{\gamma} \over {\nu}}$. Here the estimation of the exponent ${{\gamma} \over {\nu}} = 1.1887\pm 0.1645$. The goodness of fit has been
analyzed by the standard $\chi^2$ test. The value of the $\chi^2$ for three degrees of
freedom (here we have five data points and two fitting parameters) is 2.206 which
is much less than the value (7.810) of that of five percent significant level. 

\begin{figure}[h!]
\centering
	\includegraphics[angle=0,width=\textwidth]{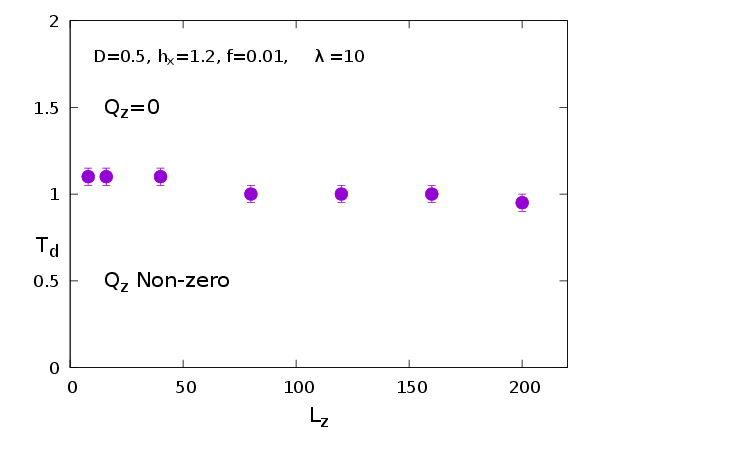}
\caption{The nonequilibrium phase transition temperature $T_d$ is plotted against
the different longitudinal sizes ($L_z$) of the nanorod.}
\label{size}
\end{figure}

We have also studied the size ($L_z$) dependence of the dynamic transition temperature
($T_d$) and shown in Fig-\ref{size}. Here, the dynamic transition temperatures for
different lengths ($L_z$) of the nanorod are studied. The plot of $T_d$ against the $L_z$ does not show any remarkable dependence of the transition temperature ($T_d$) on the length of the nanorod $L_z$.

\vskip 2 cm
\noindent {\large\bf V. Summary}

The tiny anisotropic Heisenberg ferromagnet (rodlike) in the presence
of propagating magnetic field wave has been studied by Monte Carlo simulation using
Metropolis single spin-flip algorithm. The direction of propagation of the magnetic wave has been considered here as the z-direction (which is an easy anisotropy axis here).
The z-component of the time-averaged magnetization over a complete cycle of the propagating field was found to assume nonzero value at any finite temperature if the system is cooled down from high temperature. This acts as the order parameter of such a nonequilibrium phase
transition. This is an indication of the nonequilibrium phase transition. The transition temperature was found from the temperature dependence of the variance of
the component of the dynamic order parameter. The dynamic transition was found to
occur at a lower temperature for the higher value of the amplitude of the propagating
magnetic field wave. A comprehensive phase boundary was drawn. The phase boundary does
not show any significant deviation if the wavelength of the propagating wave changes. But notable changes of the
phase boundary were observed as the strength of anisotropy changes. The region of the ordered phase was found to expand for a stronger value of anisotropy.

The size dependence of the variance of the component
of the dynamic order parameter shows the growth of dynamic correlation in the vicinity of the dynamic transition temperature. The size dependence of the transition temperature
($T_d$) does not show any significant changes within the domain of length of the
nanorod considered in this study.

One may think of reducing the size of the cross-sectional area of the nanorod to
make the system quasi-one-dimensional. A question may arise whether the
nonequilibrium phase transition can be observed in such a quasi-one-dimensional system. However, we have to keep in mind the following two important issues in this context. (i) As pointed out by the Mermin-Wagner theorem that classical Heisenberg ferromagnet can show long-range ferromagnetic ordering (Ferro-para
phase transition) only above two dimensions and this is an equilibrium concept. (ii) Here the transition studied is a nonequilibrium type which
is an outcome of the competition between the intrinsic timescale
(relaxation time) of the system and the tunable (time period) timescale
of the external field.
It may be mentioned here that this kind of dynamic phase transition disappears \cite{rmp} in the vanishingly small frequency limit, in all dimensions, not only in one dimension.

 \vskip 0.2 cm

\vskip 2 cm
\noindent {\large\bf V. Acknowledgements}
	
The financial support through FRPDF, Presidency University is gratefully acknowledged.

\vskip 0.2 cm


\newpage


\newpage


\newpage

\newpage

\newpage


\newpage


\newpage


\vskip 2 cm
\begin{thebibliography}{99}

\bibitem{rana} Bivas Rana, Amrit Kumar Mondal, Supriyo Bandyopadhyay and Anjan Barman,
arxiv:2102.12436.
\bibitem{irshad}M.I. Irshad, F. Ahmad and N.M. Mohamed,  AIP Conf. Proc., vol. 1482, (2012) 625.
\bibitem{barman} S. Bedanta, A. Barman, W. Kleemann, O. Petracic, and T. Seki,
Journal of Nanomaterials, 2013 (2013) 952540

\bibitem{erol2} E. Vatansever and M. Acharyya, J. Magn. Magn. Mater., 527 (2021) 167721
\bibitem{kantar1}E. Kantar, B. Deviren and M. Keskin,  Eur. Phys. J. B., 86, (2013) 253.
\bibitem{yuksel1} Y. Yuksel, Phys. Rev. E, 91 (2015) 032149
\bibitem{yuksel2} Y. Yüksel,  J. Magn. Magn. Mater., 389, (2015) 34.

\bibitem{erol1} E. Vatansever, Academic Platform Journal of Engineering and Science 6, (2018) 72
\bibitem{deviren2}B. Deviren, Y. Sener, M. Keskin,   392, (2013) 3983.
\bibitem{ertas1}M. Ertaş and Y. Kocakaplan,  Phys. Lett. A, vol. 378, (2014) 845.
\bibitem{keskin1}B. Deviren, E. Kantar and M. Keskin,  J. Magn. Magn.
Mater., 324, (2012) 2163.

\bibitem{binder}K. Binder, Monte Carlo Methods in Statistical
Physics, Springer, Berlin, 1979.
\bibitem{newman} M.E.J. Newman and G.T. Barkema, Monte Carlo
Methods in Statistical Physics, Clarendon Press, Oxford,
2001.

\bibitem{ma1} M. Acharyya, Phys. Rev. E, 69 (2004) 027105
\bibitem{ma2} M. Acharyya, Int. J. Mod. Phys. C, 17 (2006) 1107
\bibitem{robb} D.T. Robb, Y.H. Xu, O. Hellwig, J. McCord, A.
Berger, M.A. Novotny and P.A. Rikvold,  Phys. Rev. B., 78, (2008) 134422.
\bibitem{berger1}M. Quintana and A. Berger, Phys. Rev. E, 104 (2021) 044125
\bibitem{berger2}A. Berger, O. Idigoras and P. Vavassori,  Phys. Rev. Lett., 111, (2013) 190602.
\bibitem{berger3}P. Riego, P. Vavassori and A. Berger,  Phys. Rev.
Lett., 118, (2017) 117202.
\bibitem{rmp} B. K. Chakrabarti and M. Acharyya, Rev. Mod. Phys. 71 (1999) 847

\end{thebibliography}
\end{document}